\documentclass[reprint,groupedaddress,amsmath,amssymb,aps,prl,floatfix]{revtex4-1}

\usepackage{amsmath,amssymb,graphicx,mathrsfs,amsfonts}

\draft 

\begin{document}

\title{Strong Coupling on a Forbidden Transition in Strontium and Nondestructive Atom Counting} 



\author{Matthew A. Norcia}
\affiliation{JILA, NIST, and University of Colorado, 440 UCB, 
Boulder, CO  80309, USA}
\author{James K. Thompson}
\affiliation{JILA, NIST, and University of Colorado, 440 UCB, 
Boulder, CO  80309, USA}
\email[]{kevin.cox@jila.colorado.edu}

\date{\today}

\begin{abstract}
We observe strong collective coupling between an optical cavity and the forbidden spin singlet to triplet optical transition  $^1$S$_0$ to $^3$P$_1$  in an ensemble of $^{88}$Sr. Despite the transition being 1000 times weaker than a typical dipole transition, we observe a well resolved vacuum Rabi splitting. We use the observed vacuum Rabi splitting to make non-destructive measurements of atomic population with the equivalent of projection-noise limited sensitivity and minimal heating ($<0.01$ photon recoils/atom).  This technique may be used to enhance the performance of optical lattice clocks by generating entangled states and reducing dead time.

\end{abstract}


\maketitle

Two modes are strongly coupled when the frequency  $\Omega$ at which they exchange excitations exceeds the rate of interaction with the environment.   Strong coupling enables one to generate large amounts of entanglement \cite{Squeezing_Bohnet_2014,PhysRevA.89.043837}, achieve efficient quantum memories \cite{PhysRevLett.98.183601}, cool the motion of atoms \cite{PhysRevLett.107.143005} and mesoscopic oscillators \cite{Lehnert_Cooling_11}, and explore collective self-organization and synchronization phenomena such as superradiant lasing \cite{BCW12,2015arXiv150306464W,PhysRevA.90.053845,PhysRevLett.113.154101} and the Dicke phase transition \cite{PhysRevLett.91.203001,BGB2010}.

In this Letter, we observe strong coupling between an optical cavity and the collective excitation of up to $N=1.25\times 10^5$ strontium atoms in a 1D optical lattice.  The strong coupling is achieved using the forbidden electronic spin singlet to triplet optical transition $^1$S$_0$ to $^3$P$_1$ in $^{88}$Sr.  The excited state $^3$P$_1$ couples to the environment via spontaneous emission at the relatively slow rate of $\gamma= 2\pi \times 7.5~$kHz.

\begin{figure}[!htb]
\includegraphics[width=3.375in]{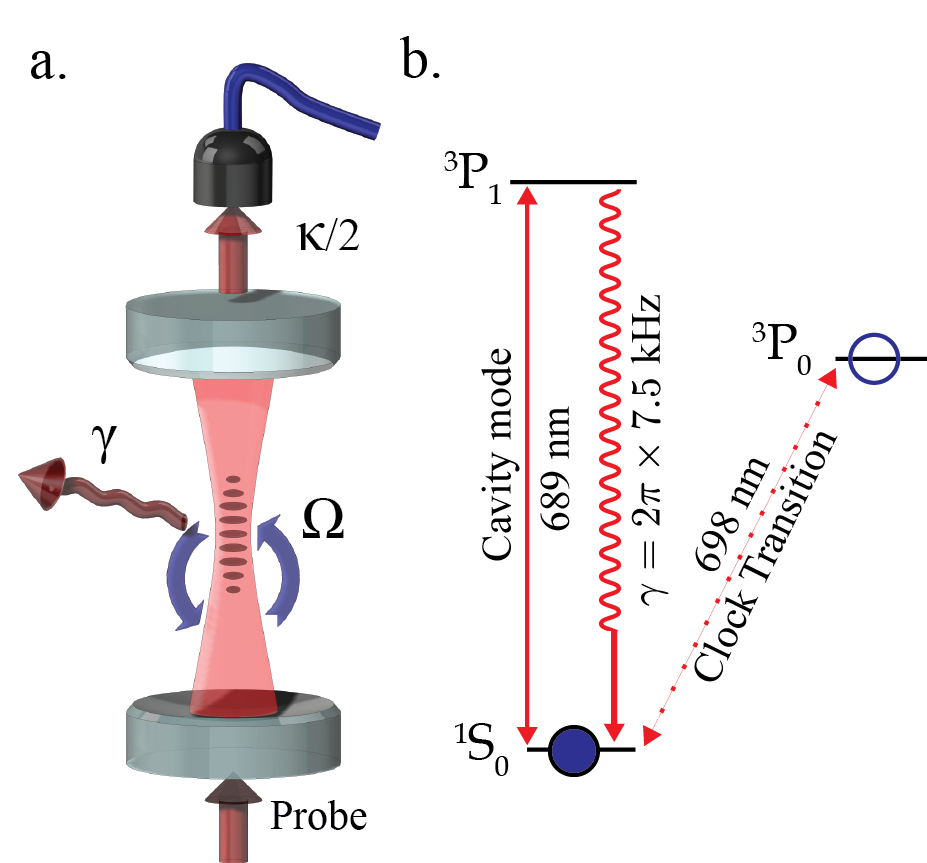}
\caption{(a) Experimental diagram.  $^{88}$Sr atoms are confined by 1D optical lattice in high-finesse cavity.  Probe light is coupled through cavity and detected on a single-photon counting module.  (b) Relevant level structure of $^{88}$Sr, showing dipole forbidden transitions at 689~nm and 698~nm.  }
\label{fig:ExpDiagram}
\end{figure}

Despite operating on a transition with 1000 times smaller squared matrix element than transitions typically used in optical cavity-QED experiments, we observe a highly-resolved splitting of the normal modes of the coupled atom-cavity system, known as a collective vacuum Rabi splitting \cite{Raizen_norm_modes, PhysRevLett.64.2499}. This observation demonstrates that despite the relative feebleness of the transition, the tools of cavity-QED can now be applied to a system of extreme interest for quantum metrology \cite{Katori2011}.  Example technologies include optical lattice clocks  \cite{Hinkley2013,Bloom14,Katori2015} and ultra-narrow lasers \cite{BCW12,2015arXiv150306464W,PhysRevA.90.053845,PhysRevLett.113.154101}, along with their associated broad range of potential applications such defining the second \cite{Katori2011, le2013experimental}, quantum many-body simulations \cite{Zhang19092014}, measuring gravitational potentials \cite{Bjerhammar1985} and gravity waves \cite{2015arXiv150100996L}, and searches for physics beyond the standard model \cite{Rosenband28032008,PhysRevLett.100.140801}.

One relevant application of this newly achieved regime is for state-selective, non-destructive counting of strontium atoms.  Such counting has been used to generate highly spin-squeezed states  \cite{Squeezing_Bohnet_2014,CWG15} that surpass the standard quantum limit on phase estimation \cite{AWO09,SLV10,CBS11}.  More simply, but quite importantly, non-destructive readout methods \cite{LWL09} can reduce the highly deleterious effects of local oscillator noise aliasing \cite{Westergaard2010,Dick1987} in optical lattice clocks.

Here, we utilize the observed vacuum Rabi splitting to non-destructively count atoms with the equivalent of sub-projection noise sensitivity.  We verify that this sensitivity is achieved with as few as 0.01 photon recoils  imparted to each atom, a well defined criteria for non-destructiveness.  These proof-of-principle measurements pave the way for the generation of entangled squeezed states on the $^1$S$_0$ to $^3$P$_0$ optical clock transition  in Yb and Sr optical lattice clocks \cite{Hinkley2013,Bloom14,Katori2015}.

Cavity-enhanced nonlinear spectroscopy \cite{MMT11} of the same transition has been performed in a MOT in \cite{PhysRevLett.114.093002}, but inhomogeneous doppler broadening prevented the observation of a collective vacuum Rabi splitting. 
Non-fluorescence based atom-counting in $^{87}$Sr  \cite{LWL09,Westergaard2010} and $^{171}$Yb \cite{PhysRevLett.102.033601} have been performed using the dipole-allowed transition $^1$S$_0$ to $^1$P$_1$ and no optical cavity.  Atomic projection noise level sensitivity was achieved, but with the scattering of $ \sim100$ photons per atom  \cite{LWL09,Westergaard2010} and with uncharacterized scattering \cite{PhysRevLett.102.033601}.  While \cite{LWL09,Westergaard2010} demonstrated that the atoms were retained in the lattice trap, the need to re-cool the atoms leads to additional stochastic losses \cite{PhysRevA.61.045801}.  Finally, at values of  $\gtrapprox 0.5$ recoils  per atom one would be unable to generate entangled states for fundamental enhancements in measurement precision beyond the standard quantum limit.

In our system, the single-atom cooperativity parameter is $C= (2 g)^2/\kappa \gamma = 0.41(4)$, where $2 g$ is the frequency of interaction between a single atom and the cavity mode \cite{jaynes1963comparison,Kimble1998}, and $\kappa= 2\pi\times 160(16)$~kHz is the cavity power decay rate.  Importantly, the interaction frequency between $N$ atoms and the cavity mode $\Omega=\sqrt{N} 2 g$ is collectively enhanced. The system is near the single-atom strong coupling regime (i.e. $C\geq 1$), but very deep into the desired collective strong coupling regime $N C = \Omega^2/\kappa \gamma=5 \times 10^4$.

In addition to the strong coupling requirement $N C \gg 1$, a useful system must have a coupling channel by which the experimentalist can extract useful information.  The decay of photons from the optical cavity at rate $\kappa$ provides collective information about the state of the atoms.  
By contrast, the atomic decay at rate $\gamma$ is difficult to utilize, leaks single-atom information out of the system, limits the generation of entangled states, and leads to heating of the ensemble.  In this Letter,  we demonstrate operation in the desired strong-coupling bad-cavity regime with hierarchy $\Omega\gg \kappa \approx 20 \times \gamma$.  In order to achieve the same favorable ratio of $\gamma$ to $\kappa$ in a system with a dipole-allowed transition, one would have to increase the cavity linewidth by a factor of one thousand either by reducing the finesse, which is fundamentally undersirable, or the cavity length, which leads to technical challenges.

A simplified experimental diagram is shown in Fig.~\ref{fig:ExpDiagram}.   The atoms are loaded from a MOT into a 1D optical lattice with wavelength  813~nm,  supported by a TEM$_{00}$ mode of the cavity. The peak trap depth is 100(10)~$\mu$K, and the final atom temperature is 4(1)~$\mu$K.

Inhomogeneous doppler broadening is highly suppressed along the cavity axis by confining the atoms in the 1D optical lattice to much less than the 689~nm probe wavelength (Lamb-Dicke parameter of 0.16).  We estimate that inhomogeneous transition broadening from the magic wavelength lattice is at most 44~kHz rms \cite{PhysRevLett.91.053001}. Since the inhomogeneous broadening is much less than $\Omega$, its impact on the dressed modes is negligible \cite{PhysRevA.53.2711}.

At 689~nm, the cavity has a finesse of $F=2.4(2)\times10^4$.   The atom-cavity coupling is enhanced by using up to $N = 1.25\times 10^5$ atoms, such that $\Omega= \sqrt{N} 2 g = 2 \pi \times 6.3$~MHz, where $g = 2 \pi \times 9.2(1)$~kHz \cite{jaynes1963comparison,Kimble1998}.  The quoted $N$ and $g$ are the effective atom numbers and couplings used to account for inhomogeneous coupling of atoms to the standing wave probe mode  \cite{geff_ref_note,SLV10,CBS11}.

\begin{figure}[!thb]
\includegraphics[width=3.375in]{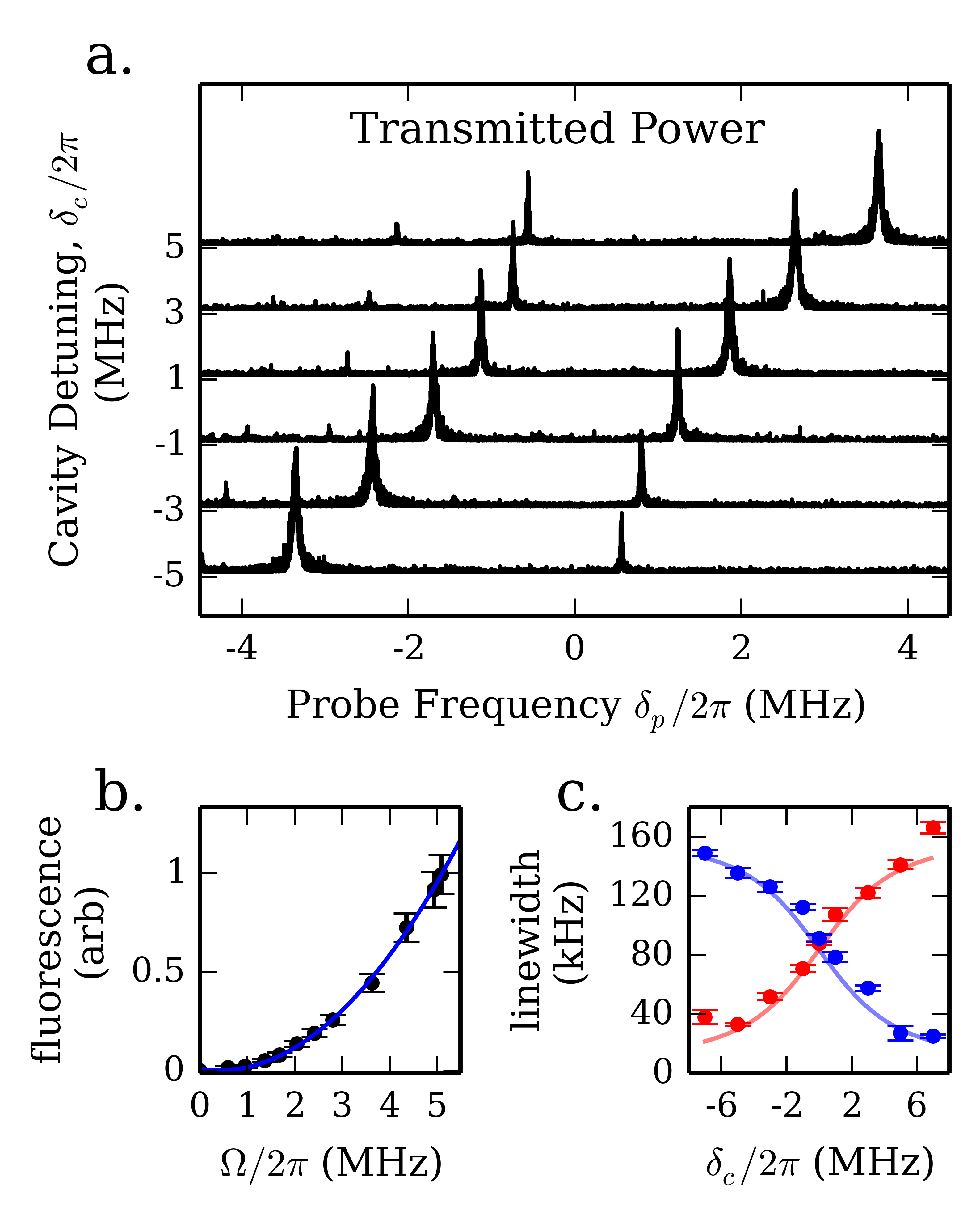}
\caption{Observation of the collective strong coupling regime on the forbidden $^1$S$_0$ to $^3$P$_1$ transition.  (a) We characterize the vacuum Rabi Splitting by recording transmitted power as probe frequency is swept.  Traces correspond to different cavity detunings, $\delta_c$. (b) Atom number, as inferred from fluorescence detection, versus vacuum Rabi splitting $\Omega$.  (c) Fitted linewidths $\kappa'_\pm$ of the two resonances at $\omega_+$ (red) and $\omega_-$ (blue) versus $\delta_c$, with prediction based on known atomic linewidth $\gamma$, measured $\Omega$ and cavity linewidth $\kappa$.  }
\label{fig:VRS}
\end{figure}

Figure \ref{fig:VRS}a shows a vacuum Rabi splitting obtained by tuning the empty cavity resonance $\omega_c$ near the atomic transition frequency $\omega_a$, loading atoms into the lattice, and then recording the transmitted power as a cavity probe is linearly swept in frequency.  The two transmission peaks correspond to the new normal modes of the dressed atom-cavity system at frequencies $\omega_+$ and $\omega_-$.  The different traces correspond to different detunings $\delta_c = \omega_c - \omega_a$, and exhibit the expected avoided-crossing behavior \cite{Raizen_norm_modes}.  In Fig.~\ref{fig:VRS}b, we confirm the $\sqrt{N}$ scaling of the size of the normal mode splitting with atom number by measuring the frequency splitting of the two modes at zero detuning, and comparing to the subsequently measured free-space fluorescence signal when 461~nm light is applied from the sides of the cavity.


The fitted widths $\kappa'_\pm$ of the upper and lower transmission peaks are plotted in Fig.~\ref{fig:VRS}b, along with a theoretical prediction based on the known atomic decay rate $\gamma$ and the measured cavity linewidth $\kappa$ \cite{Raizen_norm_modes}.  This linewidth averaging further confirms the strong coupling between atoms and cavity.

To demonstrate the ability to achieve non-destructive readout on a forbidden transition, we mimic the measurement sequences used elsewhere to achieve 10 to 14~dB of observed spin squeezing on an allowed dipole transition in $^{87}$Rb \cite{Squeezing_Bohnet_2014,CWG15}. 
We characterize the non-destructiveness of the measurement by the average number of photon recoils  per atom $m_{s}$ imparted while achieving a measurement imprecision equal to the  projection noise level of the atoms, $m_s^{QPN}$ \cite{Wineland1992,Itano1993}.  In our system, the atoms are not in a superposition of states, so there is no atomic projection noise.  However, to give a point of comparison, we define the projection noise level as though our $N$ atoms were actually $2N$ atoms, each in a superposition of the ground state and the excited $^3$P$_0$ clock state, represented in Fig.~\ref{fig:ExpDiagram}b.  The rms quantum projection noise in the ground state population would then be $\Delta N_{PN}= \sqrt{N/2}$.   We consider the rms noise $\Delta N=\Delta(N_f-N_p)$ in the difference of two consecutive measurements $N_p$ and $N_f$ of the population in the ground state $^1$S$_0$.  We define the spin noise reduction as $R= (\Delta N / \Delta N_{PN})^2.$  Two figures of merit for the measurement are the minimum $R$ attainable, and $m_s^{QPN}$.

Two consecutive measurements of the vacuum Rabi splitting $\Omega$, with measurement outcomes labeled $\Omega_p$ and $\Omega_f$, are used to infer the populations of the ground state $N_p$ and $N_f$ \cite{CBS11}.
For $\delta_c=0$, the rms fluctuations $\Delta N_{PN}$ lead to rms fluctuations in the measured vacuum Rabi splitting  $\Delta \Omega_{PN} = g/(2\sqrt{2})=2 \pi \times 3.24(3)$ kHz, regardless of atom number $N$, provided $N$ is sufficiently large \cite{CBS11,PhysRevA.89.043837}.  The spin noise reduction can then be written in terms of measured and known quantities as $R= (\Delta(\Omega_f-\Omega_p)/\Delta \Omega_{PN})^2$.  The vacuum Rabi splitting frequency $\Omega=\omega_+-\omega_-$ is determined by measuring the transmitted power when probe light is scanned over cavity resonance.  

To gain common mode rejection of frequency noise between the cavity and probe laser, we use two frequency components or tones to simultaneously probe the two dressed cavity modes, as shown in Fig. 3a \cite{CBS11,PhysRevLett.106.133601}.  
We create the two probe frequencies $\omega_{p\pm}$ by amplitude modulating a fixed probe laser whose center frequency is resonant with the empty cavity.  
As we ramp the modulation frequency $\omega_m$ over 1~MHz in 40~ms, the lower and higher frequency AM sidebands $\omega_{p\pm}$  simultaneously sweep across the resonances at $\omega_\pm$.

When the probe tones are on the side of the resonance feature of the dressed modes, changes in the number of atoms in $^1$S$_0$ cause first order changes in the total transmitted power.  In the small noise limit, relative frequency noise between the probe laser and the empty cavity causes no net change in the total transmitted power.   To demonstrate the noise suppression of this two-tone technique, Fig. 3b shows the measured power spectral density of cavity-probe frequency noise, using a single sideband to probe the bare cavity resonance versus using the two tones $\omega_{p\pm}$ simultaneously.

To demonstrate the probing method with atoms, we perform sweeps as described above and fit the total transmitted power to a Lorentzian.  We then extract the portion of the time data that corresponds to probing on side of resonance, defined as the portion between .5 and .75 $\times \kappa/2$.  To convert changes in transmitted power into changes in frequency, we fit a linear slope $S$ of transmitted power versus the known probe frequency.  The time data is broken into bins of length $\tau$, and the average transmitted power, $P_i$ is computed within each time bin $i$.    Differences are taken between adjacent bins, and the desired standard deviation $\Delta\Omega=\Delta P/S = \Delta(P_{i+1} - P_i)/S$ is computed over non-overlapping pairs of bins.

\begin{figure}[!htb]
\includegraphics[width=3.375in]{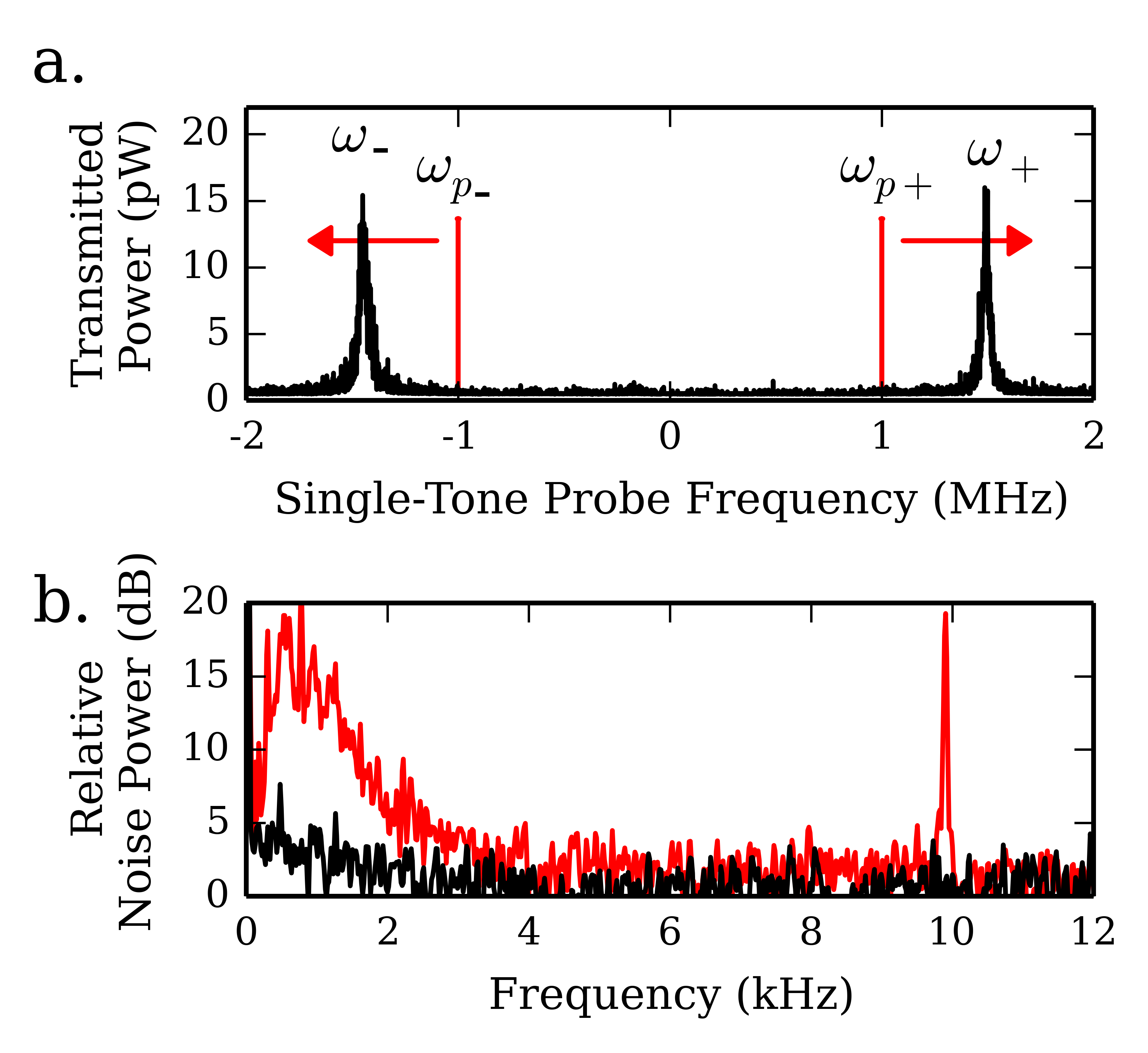}
\caption{(a) To probe the Vacuum Rabi Splitting, two probe tones at $\omega_{p\pm}$ are simultaneously swept across the two normal mode resonances at $\omega_{\pm}$ while total transmitted power is recorded.  (b) The noise power spectra for cavity probe with a single tone (red trace) and two tones (black trace), demonstrate noise cancellation of two-tone probing.}
\label{fig:fig3}
\end{figure}

Figure 4a shows the spin noise reduction $R$ as a function of the chosen window length $\tau$ at $N= 9\times 10^4$.  Similar results were obtained for atom numbers between $N= 1.5\times 10^4$ and  $1.25\times 10^5$.   At a given atom number,  the incident probe power is held fixed at a value that does not saturate the detector or the atomic ensemble.  Therefore, increasing $\tau$ is equivalent to collecting more probe photons to average down photon shot noise.  At large $\tau$, $R$ rises due to uncanceled low frequency probe noise.  For window lengths between  50 and 175~$\mu$s, corresponding to detected photon numbers of 750 and 2650 per window, we are able to measure spin noise reductions $R<1$.  The lowest value measured is $R = 0.58(13)$, with no corrections applied for noise in the second measurement window.

\begin{figure}[!htb]
\includegraphics[width=3.375in, ]{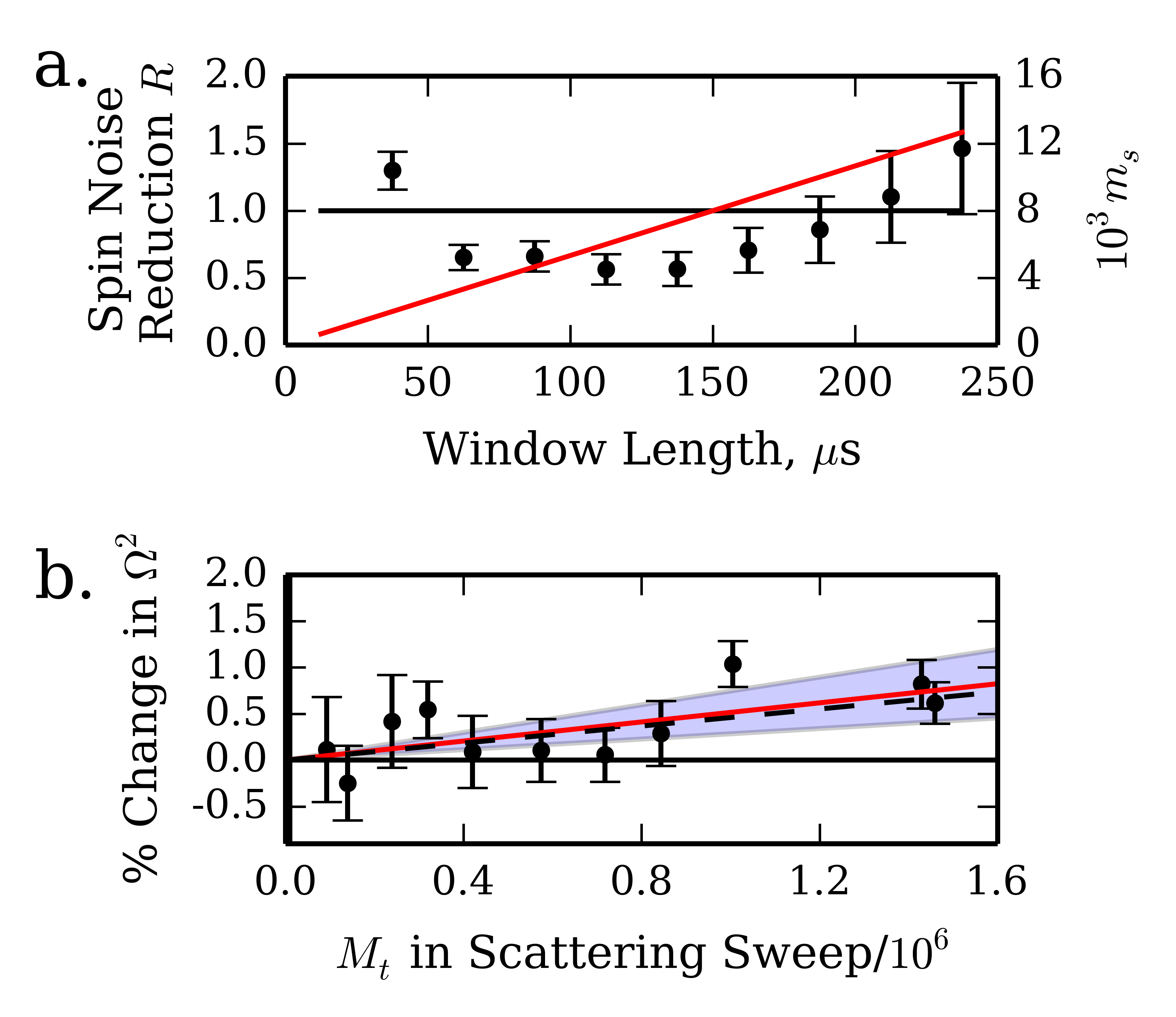}
\caption{(a) Black points show measurement noise relative to hypothetical projection noise in adjacent windows $R$, versus length of a measurement window.  The lowest two points represent $R = -2.4^{.9}_{1.1}$ dB. Red line shows the the number of scattered photons per atom $m_s$ in a single measurement window.  (b) Fractional change in $\Omega^2 = 4 g^2 N$ versus $M_t$, the number of photons in additional scattering sweep.  Solid red line is fit to data, with shaded region representing 1 sigma error on the fit.  Dashed line represents a prediction based on the free space scattering.  The number of photons used in two typical 100$~\mu $s windows is represented by the thickness of the left axis.  }
\label{fig:RandHeating}
\end{figure}

The goal of being in the bad-cavity, strongly-coupled regime is to achieve a high ratio of detected photons to free-space scattered photons, which lead to atomic state collapse, heating, and potentially atom loss.  We characterize these effects by measuring how passing photons through the cavity leads to changes in $\Omega$, as shown in Fig.~\ref{fig:RandHeating}b.  Changes in $\Omega$ allow us to infer changes in the product $g\sqrt{N}$.  These changes could either be due to atom loss or heating in directions perpendicular to the cavity axis.  We make our observations at  much higher transmitted probe photon numbers $M_t$ than those used for observing $R<1$ by inserting a third ``scattering sweep" between two probe sweeps used to measure a change in $\Omega$.   From the fit to the change in $\Omega$ vs $M_t$,  we extrapolate to the typical number of transmitted photons $M_t\approx 5000$ in a single window that achieves $R\le1$.  At $N=5\times10^4$ atoms, we find 0.01 photons scattered per atom/measurement window.  This corresponds to a temperature increase for two windows of 8~nK, or in the context of generating entanglement, a loss of atomic coherence of only 1\% or 0.09~dB of signal loss due to the first window.

We theoretically expect one free-space scattered photon per 11(1) photons transmitted from a single end of the cavity, a value set simply by the ratio $\kappa/2\gamma$.   The resulting change in $\Omega$ predicted from this scattering is shown as a dashed line in Fig.~\ref{fig:RandHeating}b, and is consistent within the error of the fit to the data.

Because we operate with $\kappa\gg\gamma$ there is little fundamental loss for generating atomic squeezing or for non-destructive readout from probing with $\delta_c = 0$, and applying the noise-immune two-tone technique. This would not be true for the current cavity geometry and finesse on a fully allowed dipole transition as the magnitudes of the atomic damping and cavity damping would be reversed $\kappa\ll\gamma$.

For certain applications, including spin squeezing, another relevant metric for the non-destructiveness of a measurement is the degree to which it imparts inhomogeneous AC Stark or light shifts of the ground state $^1$S$_0$.
While we cannot measure this directly, we can estimate it by considering the AC Stark shifts associated with the two probe tones. 
For typical probing parameters, the time integrated phase shift contributed by a single probe tone during a measurement window would be of order $\pi$ radians for maximally coupled atoms. 
With our two-tone probing scheme, the phase shifts cancel due to their opposite detunings from the excited state $^3$P$_1$ \cite{PhysRevA.89.043837}.  If we assume a power balance at the 10\% level, we could expect a phase shift of order 0.1 radians for maximally coupled atoms, such that the loss of atomic contrast would be $<0.5\%$ or 0.04~dB loss of signal.  This cancellation would potentially eliminate the need for a spin-echo sequence to maintain contrast in an experimental sequence.  Cavity-optomechanical effects that can limit the observed spin noise reduction would be partially canceled as well \cite{Squeezing_Bohnet_2014}.

From $m_s^{QPN}$, we can estimate the fundamental limit of our probing scheme to surpass the standard quantum limit on phase estimation \cite{PhysRevA.89.043837}.  
If technical sources of noise were addressed, one could achieve 22~dB of spin squeezing at $2 \times 10^5$ total atoms.
This fundamental limit on spectroscopic enhancement could reach 30~dB by improving the net effective quantum efficiency for detecting the probe light from the current 3.5\% to a conservative value of 25\%.


In conclusion, we have achieved  strong collective coupling between a cavity mode and an ensemble of atoms on a dipole-forbidden transition.  
This may enable future advances in quantum metrology, in non-linear photon interactions \cite{MEH10,PhysRevLett.67.1727,PhysRevLett.75.4710}, and in the generation of entanglement for advancing optical lattice clocks \cite{Hinkley2013,Bloom14,Katori2015}. Further, by strongly coupling a narrow transition  to a more highly damped optical cavity ($\kappa \gg \gamma$), the rate at which information can be extracted is greatly increased.  This may open the door to clock or qubit readout in systems for which strong dipole-allowed transitions do not exist at accessible wavelengths \cite{Schmidt29072005}.

The authors acknowledge Karl H. Mayer and Matthew N. Winchester for contributions to the experimental apparatus, as well as Jun Ye, Travis Nicholson and Sara Campbell for strontium advice, and Wei Zhang for providing reference light.  
All authors acknowledge financial support from DARPA QuASAR, ARO, NSF PFC, and NIST. This work is supported by the National Science Foundation under Grant Number 1125844.

\bibliographystyle{apsrev4-1}
\bibliography{ThompsonLab.bib}

\begin{thebibliography}{46}
\bibitem [{gef()}]{geff_ref_note}%
  \BibitemOpen
  \href@noop {} {\ }\bibinfo {note} {The effective coupling is related to the
  maximum possible coupling $g_0$ by $g= \sqrt{3/4} g_0$. The effective atom
  number is related to the total number of atoms in the lattice
  $N_{\mathrm{tot}}$ by $N= 2N_{\mathrm{tot}}/3$.}\BibitemShut {Stop}%
\end{thebibliography}

\end{document}